\documentclass[11pt]{article}%
\usepackage{amsmath}
\usepackage{amssymb}
\usepackage{amsfonts}

\usepackage{cite}

\usepackage{graphicx}
\usepackage{float}
\usepackage{longtable}
\usepackage[utf8]{inputenc}
\usepackage{hyperref}

\usepackage{algorithmicx}
\usepackage[ruled,vlined]{algorithm2e}
\SetAlgoCaptionSeparator{ }

\usepackage{fullpage}

\usepackage{authblk}

\newcommand{\fref}[1]{Fig.~\ref{#1}}

\newcommand{\tref}[1]{Table~\ref{#1}}
\newcommand{\sref}[1]{Section~\ref{#1}}

\newenvironment{prot}[1][!htbp]
  {
   \begin{algorithm}[#1]%
  }{\end{algorithm}}

\providecommand{\U}[1]{\protect\rule{.1in}{.1in}}

\begin{document}

\title{Entanglement Availability Differentiation Service for the Quantum Internet}
\author[1,2,3]{Laszlo Gyongyosi\footnote{Email: \href{mailto:l.gyongyosi@soton.ac.uk}{l.gyongyosi@soton.ac.uk}}}
\author[2]{Sandor Imre}
\affil[1]{School of Electronics and Computer Science, University of Southampton, Southampton, SO17 1BJ, UK}
\affil[2]{Department of Networked Systems and Services, Budapest University of Technology and Economics, Budapest, H-1117 Hungary}
\affil[3]{MTA-BME Information Systems Research Group, Hungarian Academy of Sciences, Budapest, H-1051 Hungary}
\date{}

\maketitle
\begin{abstract}
A fundamental concept of the quantum Internet is quantum entanglement. In a quantum Internet scenario where the legal users of the network have different priority levels or where a differentiation of entanglement availability between the users is a necessity, an entanglement availability service is essential. Here we define the entanglement availability differentiation (EAD) service for the quantum Internet. In the proposed EAD framework, the differentiation is either made in the amount of entanglement with respect to the relative entropy of entanglement associated with the legal users, or in the time domain with respect to the amount of time that is required to establish a maximally entangled system between the legal parties. The framework provides an efficient and easily-implementable solution for the differentiation of entanglement availability in experimental quantum networking scenarios.
\end{abstract}

\section{Introduction}
\label{sec1}
In the quantum Internet \cite{r2, r23}, one of the most important tasks is to establish entanglement \cite{r1, r2, r3, r4, r5, r6, r7, r8, r9, r10} between the legal parties \cite{r11, r12, r13, r14, r15} so as to allow quantum communication beyond the fundamental limits of point-to-point connections \cite{r41,r42,r43}. For the problem of entanglement distribution in quantum repeater networks several methods \cite{r7, r10, r11, r12, r13, r14, r15}, and physical approaches have been introduced \cite{r16, r17, r18, r19, r20, r21, r22, r23, r24, r25, r26, r27, r28, r29, r30, r31, r32, r33, r34, r35, r36, r37, r38, r39, r40, r47,r48,r49}. The current results are mainly focusing on the physical-layer of the quantum transmission \cite{r5, r6, r7, r8, r9}, implementations of entanglement swapping and purification, or on the optimization of quantum memories and quantum error correction in the repeater nodes \cite{r16, r17, r18, r19, r20, r21, r22, r23, r24, r25, r26, r27, r28, r29, r30, r31, r32, r33, r34, r35, r36, r37, r38, r39, r40}. However, if the legal users of the quantum network are associated with different priority levels, or if a differentiation of entanglement availability between the users is a necessity in a multiuser quantum network, then an efficient and easily implementable entanglement availability service is essential. 

In this work, we define the \textit{entanglement availability differentiation} (EAD) service for the quantum Internet. We introduce differentiation methods, Protocols 1 and 2, within the EAD framework. In Protocol 1, the differentiation is made in the amount of entanglement associated with the legal users. The metric used for the quantization of entanglement is the relative entropy of entanglement function \cite{r44,r45,r46}. In Protocol 2, the differentiation is made in the amount of time that is required to establish a maximally entangled system between the legal parties. 

The EAD framework contains a classical phase (Phase 1) for the distribution of timing information between the users of the quantum network. Phase 2 consists of all quantum transmission and unitary operations. In Phase 2, the entanglement establishment is also performed between the parties according to the selected differentiation method. 

The entanglement distribution phase of EAD utilizes Hamiltonian dynamics, which allows very efficient practical implementation for both the entanglement establishment and the differentiation of entanglement availability. Using the Hamiltonian dynamics approach as a core protocol of Step 2 of the EAD framework, the entanglement differentiation method requires only unitary operations at the transmitter and requires no entanglement transmission. The application time of the unitaries can be selected as arbitrarily small in the transmitter to achieve an efficient practical realization. The proposed EAD framework is particularly convenient for experimental quantum networking scenarios, quantum communication networks, and future quantum internet. 

The novel contributions of our manuscript are as follows:
\begin{itemize}
\item \textit{We define the entanglement availability differentiation (EAD) service for the quantum Internet.}
\item \textit{The entanglement availability differentiation is achieved via Hamiltonian dynamics between the users of the quantum network.}
\item \textit{The EAD framework can differentiate in the amount of entanglement with respect to the relative entropy of entanglement associated to the legal users (Protocol 1), and also in the time domain with respect to the amount of time that is required to establish a maximally entangled system (Protocol 2) between the legal parties.}
\item \textit{The framework provides an efficient and easily-implementable solution for the differentiation of entanglement availability in experimental quantum networking scenarios}.  
\end{itemize}

This paper is organized as follows. \sref{sec2} defines the framework for the proposed entanglement differentiation methods. \sref{sec3} discusses the entanglement differentiation schemes. Finally, \sref{sec4} concludes the results. Supplemental information is included in the Appendix.

\section{System Model}
\label{sec2}
The proposed EAD service allows differentiation in the amount of entanglement shared between the users or the amount of time required for the establishment of maximally entangled states between the users. The defined service requires no entanglement transmission to generate entanglement between the legal parties. The differentiation service consists of two phases: a classical transmission phase (Phase 1) to distribute side information for the entanglement differentiation and a quantum transmission phase (Phase 2), which covers the transmission of unentangled systems between the users and the application of local unitary operations to generate entanglement between the parties. 

The proposed entanglement availability differentiation methods are detailed in Protocol 1 and Protocol 2. The protocols are based on a core protocol (Protocol 0) that utilizes Hamiltonian dynamics for entanglement distribution in quantum communication networks (see \sref{corep}). The aim of the proposed entanglement differentiation protocols (Protocol 1 and Protocol 2) is different from the aim of the core protocol, since Protocol 0 serves only the purpose of entanglement distribution, and allows no entanglement differentiation in a multiuser quantum network. Protocol 0 is used only in the quantum transmission phase and has no any relation with a classical communication phase.

\subsection{Classical Transmission Phase}

In the classical transmission phase (Phase 1), the timing information of the local Hamiltonian operators are distributed among the legal parties by an ${\mathcal{E}}$ encoder unit. The content of the timing information depends on the type of entanglement differentiation method. The Hamiltonian operators will be applied in the quantum transmission phase (Phase 2) to generate entangled systems between the users. Since each types of entanglement differentiation requires the distribution of different timing information between the users, the distribution of classical timing information will be discussed in detail in \sref{sec3}.

\subsection{Quantum Transmission Phase}

The quantum transmission phase (Phase 2) utilizes a core protocol for the entanglement distribution protocol of the EAD framework. The core protocol requires no entanglement transmission for the entanglement generation, only the transmission of an unentangled quantum system (i.e., separable state \cite{r10, r11, r12, r13, r14, r15}) and the application of a unitary operation for a well-defined time in the transmit user. The core protocol of the quantum transmission phase for a user-pair is summarized in Protocol 0. It assumes the use of redundant quantum parity code \cite{r7} for the encoding\footnote{Actual coding scheme can be different.}. For a detailed description of Protocol 0, see \sref{corep}. 

\setcounter{algocf}{-1}
\begin{prot}
  \DontPrintSemicolon
\caption{Core Protocol}
\textbf{Step 1.} Alice (transmitter node) and Bob (receiver node) agree on a time $t$, in which they want to establish entanglement between subsystems $A$ and $B$. Alice generates a separable initial system $AB$, with no entanglement between $A$ and $B$ as 
\begin{equation} \label{eq1} 
{\rho }_{AB}=\frac{1}{2}\left|{\psi }_+\right\rangle \left\langle \left.{\psi }_+\right|\right.+\frac{1}{2}\left|{\phi }_+\right\rangle \left\langle \left.{\phi }_+\right|\right., 
\end{equation} 
 where $\left|{\psi }_+\right\rangle =\frac{1}{\sqrt{2}}\left(\left|01\right\rangle +\left|10\right\rangle \right)$, $\left|{\phi }_+\right\rangle =\frac{1}{\sqrt{2}}\left(\left|00\right\rangle +\left|11\right\rangle \right)$. Alice encodes subsystem $B$, $\left|{\varphi }_B\right\rangle =\alpha \left|0\right\rangle +\beta \left|1\right\rangle $ via an $\left(m,n\right)$ redundant quantum parity code as 
\begin{equation} \label{eq2} 
{\left|{\delta }_B\right\rangle }^{\left(m,n\right)}=\alpha {\left|{\chi }_+\right\rangle }^{\left(m\right)}_1\dots {\left|{\chi }_+\right\rangle }^{\left(m\right)}_n+\beta {\left|{\chi }_-\right\rangle }^{\left(m\right)}_1\dots {\left|{\chi }_-\right\rangle }^{\left(m\right)}_n, 
\end{equation} 
 where ${\left|{\chi }_{\pm }\right\rangle }^{\left(m\right)}={\left|0\right\rangle }^{\otimes m}\pm {\left|1\right\rangle }^{\otimes m}$, and sends subsystem $B$ to Bob through the network of $n$ intermediate transfer nodes. Each intermediate transfer nodes ${\mathcal{N}}_{1\dots n}$ receives and retransmits ${\left|{\delta }_B\right\rangle }^{\left(m,n\right)}$. Bob receives ${\left|{\delta }_B\right\rangle }^{\left(m,n\right)}$ and decodes it.

\textbf{Step 2.} Alice prepares a system $C$, denoted by density ${\rho }_C=\frac{1}{2}\left(I+s{\sigma }^x\right)$ which is completely uncorrelated from ${\rho }_{AB}$, where $I$ is the identity operator, ${\sigma }^x$ is the Pauli $X$ matrix, while $s$ is a constant. Alice applies a unitary $U'_{AC}$ on $A$ and $C$, which produces the initial system $ABC$ with no entanglement between $A$ and $B$ as 
\begin{equation} \label{eq3} 
{\rho }_{ABC}=U'_{AC}{\rho }_{AB}{\rho }_C{\left(U'_{AC}\right)}^{\dagger }=\frac{1}{2}\left|{\psi }_+\right\rangle \left\langle \left.{\psi }_+\right|\right.\left|+\right\rangle \left\langle \left.+\right|\right.+\frac{1}{2}\left|{\phi }_+\right\rangle \left\langle \left.{\phi }_+\right|\right.\left|-\right\rangle \left\langle \left.-\right|\right., 
\end{equation} 
 where $\left|\pm \right\rangle =\frac{1}{\sqrt{2}}\left(\left|0\right\rangle \pm \left|1\right\rangle \right)$.

\textbf{Step 3.} Alice applies the unitary $U_{AC}$ on subsystem $AC$ for a time $t$, as 
\begin{equation} \label{eq4} 
U_{AC}=\exp \left(-iH_{AC}t\right)=\cos \left(t\right)I-i\sin \left(t\right){\sigma }^x_A{\sigma }^x_C, 
\end{equation} 
 where 
\begin{equation} \label{eq5} 
H_{AC}={\sigma }^x_A{\sigma }^x_C 
\end{equation} 
 is the Hamiltonian with energy $E_{AC}$ 
\begin{equation} \label{eq6} 
{{E}_{AC}}=\tfrac{1}{2}\hbar 2\pi \left( \tfrac{1}{4t} \right), 
\end{equation} 
 where $\hbar $ is the reduced Planck constant, which results in the maximally entangled $AB$ system with probability $p=1$ as 
\begin{equation} \label{eq7} 
{\sigma }_{AB}=\frac{1}{2}\left(\left|{\psi }_+\right\rangle -i\left|{\phi }_+\right\rangle \right)\left(\left\langle \left.{\psi }_+\right|\right.+i\left\langle \left.{\phi }_+\right|\right.\right). 
\end{equation} 
\end{prot}

\subsection{Framework}

 In our multiuser framework, the quantum transmission phase is realized by the core protocol of Phase 2; however, time $t$ of the Hamiltonian operator is selected in a different way among the users, according to the selected type of differentiation. For an $i$-th user $U_i$, the application time of the local unitary is referred to as $T_{U_i}$. Without loss of generality, the $i$-th transmit user is referred to as $U_i$, and the $i$-th receiver user is $B_i$.

In the system model, the user pairs can use the same physical quantum link, therefore in the physical layer the users can communicate over the same quantum channel. On the other hand, in a logical layer representation of the protocols, the communication between the user pairs formulate logically independent channels.

\begin{center}
\begin{figure}[!h]
\begin{center}
\includegraphics[angle = 0,width=1\linewidth]{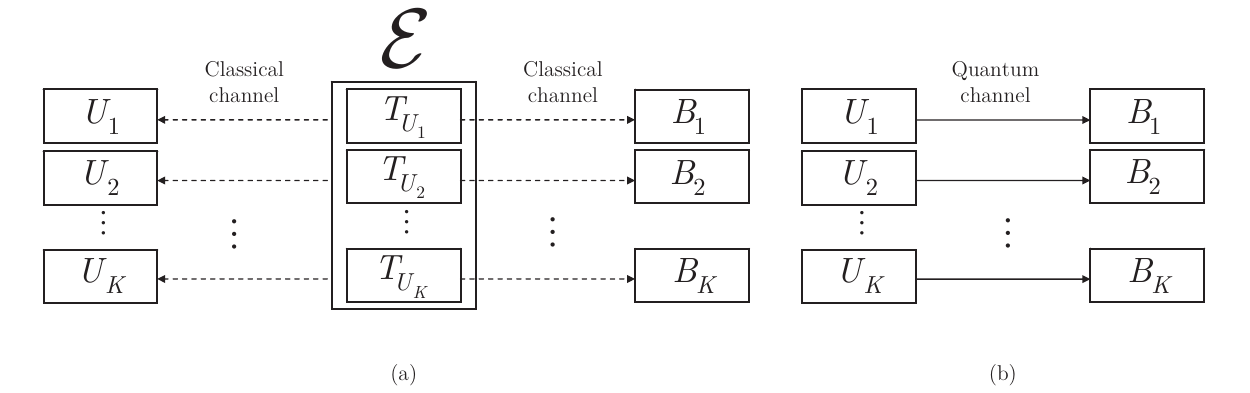}
\caption{Framework of the entanglement differentiation service in a multiuser quantum network. \textbf{(a):} Phase 1. Classical transmission. The ${\mathcal{E}}$ encoder unit distributes the timing information for the legal transmit users $U_{{\rm 1}} ,\ldots ,U_{K} $ and receiver users $B_{{\rm 1}} ,\ldots ,B_{K} $ via a classical channel. \textbf{(b):} Phase 2. Quantum transmission. The users apply the core protocol for the entanglement establishment. Then, using the received timing information the transmit users $U_{1} ,\ldots ,U_{K} $ apply the local unitaries for time $T_{U_{{\rm 1}} } {\rm ,}\ldots , T_{U_{K} } $.} 
 \label{fig1}
 \end{center}
\end{figure}
\end{center}

The method of entanglement differentiation service is summarized in \fref{fig1}. The basic model consists of two phases: distribution of timing information over classical links (\fref{fig1}(a)) and the transmission of quantum systems and the application of local unitary operations (\fref{fig1}(b)). 

\section{Methods of Entanglement Availability Differentiation}
\label{sec3}
 The EAD service defines different types of differentiation. The differentiation can be achieved in the amount of entanglement in terms of the relative entropy of entanglement between the users (\textit{Protocol 1}: differentiation in the amount of entanglement). In this method, all users have knowledge of a global oscillation period \cite{r10} of time for the application of their local unitaries, but the users will get different amounts of entanglement as a result. 

 The differentiation is also possible in the amount of time that is required to establish a maximally entangled system between the users (\textit{Protocol 2}: differentiation in the time domain). In this method, all users get a maximally entangled system as a result; however, the time that is required for the entanglement establishment is variable for the users, and there is also no global oscillation period of time.

\subsection{Differentiation in the Amount of Entanglement}

The differentiation of the entanglement amount between the users allows us to weight the entanglement amount between the users in terms of relative entropy of entanglement. Using the timing information distributed in Phase 1 between the $K$ transmit users $U_{1} ,\ldots ,U_{K} $ 
\begin{equation} \label{eq8} 
T_{U_{i} } =x_{U_{i} } +\left(\pi /{\rm 4}\right),i=1,\ldots ,K, 
\end{equation} 
 where 
\begin{equation} \label{eq9} 
x_{U_{i} } \in \left[-\left(\pi /{\rm 4}\right),\left(\pi /{\rm 4}\right)\right], 
\end{equation} 
 for an $i$-th transmit user $U_{i} $, the protocol generates an initial system $ABC$, transmits separable B to receiver $B_{i} $, and applies the local unitary $U_{AC} $ on subsystem $AC$ for time $T_{U_{i} } $ (using the core protocol of Phase 2). Depending on the selected $T_{U_{i} } $, the resulting $AB$ subsystem between users $U_{i} $ and $B_{i} $ contains the selected amount of entanglement, 
\begin{equation} \label{eq10} 
E^{\left(T_{U_{i} } \right)} \left(U_{i} :B_{i} \right)\le {\rm 1}. 
\end{equation} 
 
\subsubsection{Relative Entropy of Entanglement}

In the proposed service framework, the amount of entanglement is quantified by the $E\left(\cdot \right)$ relative entropy of entanglement function. By definition, the $E\left(\rho \right)$ relative entropy of entanglement function of a joint state $\rho $ of subsystems $A$ and $B$ is defined by the $D\left(\left. \cdot \right\| \cdot \right)$ quantum relative entropy function, without loss of generality as 
\begin{equation} \label{eq11} 
E\left( \rho  \right)=\underset{{{\rho }_{AB}}}{\mathop{\min }}\,D\left( \left. \rho  \right\|{{\rho }_{AB}} \right)=\underset{{{\rho }_{AB}}}{\mathop{\min }}\,\text{Tr}\left( \rho \log \rho  \right)-\text{Tr}\left( \rho \log \left( {{\rho }_{AB}} \right) \right), 
\end{equation} 
 where $\rho _{AB} $ is the set of separable states $\rho _{AB} =\sum _{i=1}^{n} p_{i} \rho _{A,i} \otimes \rho _{B,i} $.

\subsubsection{Differentiation Service}

 The Phases 1 and 2 of the method of entanglement amount differentiation (Protocol 1) are as included in Protocol 1. 
 
\begin{prot}
  \DontPrintSemicolon
\caption{Differentiation in the Amount of Entanglement}

\textbf{Step 1}. Let $U_{i} ,i=1,\ldots ,K$ be the set of transmit users, and $B_{i} ,i=1,\ldots ,K$ are the receiver users. Distribute the $T_{U_{i} } =x_{U_{i} } +\left(\pi /{\rm 4}\right),i=1,\ldots ,K$, where $x_{U_{i} } \in \left[-\left(\pi /{\rm 4}\right),\left(\pi /{\rm 4}\right)\right]$, timing information via an encoder unit ${\mathcal{E}}$ between all transmit users using a classical authenticated channel (Phase 1).

\textbf{Step 2}. In the transmit user $U_{i} $, generate the initial system $ABC$, transmit separable B to receiver $B_{i} $, and apply the local unitary $U_{AC} $ on subsystem $AC$ for time $T_{U_{i} } $ (Core protocol of Phase 2 between the users).

\textbf{Step 3}. The resulting $AB$ subsystem after total time $T=T_{U_{i} } $ between users $U_{i} $ and $B_{i} $ contains entanglement $E^{\left(T_{U_{i} } \right)} \left(U_{i} :B_{i} \right)=\sin ^{2} \left(2\left(\frac{\pi }{4} +x_{U_{i} } \right)\right)$.

\end{prot}

\paragraph{Description}
 In the quantum transmission phase, the entanglement oscillation in $AB$ is generated by the energy $E$ of the Hamiltonian $H$ \cite{r10}. This oscillation has a period of time $T_{\pi } $, which exactly equals to ${\rm 4}t$, 
\begin{equation} \label{eq12} 
T_{\pi } =4t, 
\end{equation} 
 where $t$ is determined by Alice and Bob. In other words, time $t$ identifies $\pi /{\rm 4}$, where $\pi $ is the oscillation period. Therefore, in Protocol 1, the density $\sigma _{ABC} $ of the final $ABC$ state is as 
\begin{equation} \label{eq13} 
\begin{split} {\sigma _{ABC} }&=  {\left| \varphi \left(t\right)\right\rangle  \left\langle \left. \varphi \left(t\right)\right|\right. _{ABC} =U\rho _{0} U^{\dag } } \\ {} & ={\frac{{\rm 1}}{{\rm 2}} \left(U_{AC} \left| \psi _{+} \right\rangle  \left\langle \left. \psi _{+} \right|\right. \left| +\right\rangle  \left\langle \left. +\right|\right. U_{AC}^{\dag } \right)+\frac{{\rm 1}}{{\rm 2}} \left(U_{AC} \left| \phi _{+} \right\rangle  \left\langle \left. \phi _{+} \right|\right. \left| -\right\rangle  \left\langle \left. -\right|\right. U_{AC}^{\dag } \right),} \end{split}
\end{equation} 
 where $\left| \varphi \left(t\right)\right\rangle  _{ABC} $ at time $t$ is evaluated as 
\begin{equation} \label{eq14} 
\begin{split} & {\left| \varphi \left(t\right)\right\rangle  _{ABC} } \\ {} & {=\frac{{\rm 1}}{\sqrt{{\rm 2}} } \left(\cos \left(t\right)\left(\left| \psi _{+} \right\rangle  \left| +\right\rangle  \right)-i\sin \left(t\right)\left(\left| \phi _{+} \right\rangle  \left| +\right\rangle  \right)\right)+\frac{{\rm 1}}{\sqrt{{\rm 2}} } \left(\cos \left(t\right)\left(\left| \phi _{+} \right\rangle  \left| -\right\rangle  \right)+i\sin \left(t\right)\left(\left| \psi _{+} \right\rangle  \left| -\right\rangle  \right)\right)} \\ {} & {=\frac{{\rm 1}}{\sqrt{{\rm 2}} } \left(\cos \left(t\right)\left(\left| \psi _{+} \right\rangle  \right)-i\sin \left(t\right)\left(\left| \phi _{+} \right\rangle  \right)\right)\left| +\right\rangle  +\frac{{\rm 1}}{\sqrt{{\rm 2}} } \left(\cos \left(t\right)\left(\left| \phi _{+} \right\rangle  \right)+i\sin \left(t\right)\left(\left| \psi _{+} \right\rangle  \right)\right)\left| -\right\rangle  } \end{split} 
\end{equation} 
 which at $T_{U_{i} } $ (see \eqref{eq8}) of user $U_{i} $, for a given $x_{U_{i} }$ is evaluated as 
\begin{equation} \label{eq17} 
 \begin{split} {\left| \varphi \left(T_{U_{i} } \right)\right\rangle  _{ABC} }  {} 
\\ {}={}&{\frac{{\rm 1}}{\sqrt{{\rm 2}} } \left(\cos \left(\frac{\pi }{{\rm 4}} +x_{U_{i} } \right)\left(\left| \psi _{+} \right\rangle  \right)-i\sin \left(\frac{\pi }{{\rm 4}} +x_{U_{i} } \right)\left(\left| \phi _{+} \right\rangle  \right)\right)\left| +\right\rangle  } 
\\&{}+{}{\frac{{\rm 1}}{\sqrt{{\rm 2}} } \left(\cos \left(\frac{\pi }{{\rm 4}} +x_{U_{i} } \right)\left(\left| \phi _{+} \right\rangle  \right)+i\sin \left(\frac{\pi }{{\rm 4}} +x_{U_{i} } \right)\left(\left| \psi _{+} \right\rangle  \right)\right)\left| -\right\rangle  } 
\\{}={}&{\frac{{\rm 1}}{\sqrt{{\rm 2}} } \left(\begin{array}{l} {\left(\frac{{\rm 1}}{\sqrt{{\rm 2}} } \left(\cos \left(x_{U_{i} } \right)\right)-\frac{{\rm 1}}{\sqrt{{\rm 2}} } \left(\sin \left(x_{U_{i} } \right)\right)\right)\left(\left| \psi _{+} \right\rangle  \right)} \\ {-i\left(\frac{{\rm 1}}{\sqrt{{\rm 2}} } \left(\cos \left(x_{U_{i} } \right)\right)+\frac{{\rm 1}}{\sqrt{{\rm 2}} } \left(\sin \left(x_{U_{i} } \right)\right)\right)\left(\left| \phi _{+} \right\rangle  \right)} \end{array}\right)\left| +\right\rangle  } \\&{}+{} {\frac{{\rm 1}}{\sqrt{{\rm 2}} } \left(\begin{array}{l} {\left(\frac{{\rm 1}}{\sqrt{{\rm 2}} } \left(\cos \left(x_{U_{i} } \right)\right)-\frac{{\rm 1}}{\sqrt{{\rm 2}} } \left(\sin \left(x_{U_{i} } \right)\right)\right)\left(\left| \phi _{+} \right\rangle  \right)} \\ {+i\left(\frac{{\rm 1}}{\sqrt{{\rm 2}} } \left(\cos \left(x_{U_{i} } \right)\right)+\frac{{\rm 1}}{\sqrt{{\rm 2}} } \left(\sin \left(x_{U_{i} } \right)\right)\right)\left(\left| \psi _{+} \right\rangle  \right)} \end{array}\right)\left| -\right\rangle  ,} \end{split} 
\end{equation} 
 where the sign change on $U_{AC} \left(\left| \phi _{+} \right\rangle  \left| -\right\rangle  \right)$ is due to the $\left| -\right\rangle  $ eigenstate on $C$, and where 
\begin{equation} \label{eq18} 
\begin{split} &{\frac{{\rm 1}}{\sqrt{{\rm 2}} } \left(\cos \left(x_{U_{i} } \right)\right)-\frac{{\rm 1}}{\sqrt{{\rm 2}} } \left(\sin \left(x_{U_{i} } \right)\right)\left(\left| \phi _{+} \right\rangle  \right)+i\left(\frac{{\rm 1}}{\sqrt{{\rm 2}} } \left(\cos \left(x_{U_{i} } \right)\right)+\frac{{\rm 1}}{\sqrt{{\rm 2}} } \left(\sin \left(x_{U_{i} } \right)\right)\right)\left(\left| \psi _{+} \right\rangle  \right)} \\&= {i\left(\left(\frac{{\rm 1}}{\sqrt{{\rm 2}} } \left(\cos \left(x_{U_{i} } \right)\right)+\frac{{\rm 1}}{\sqrt{{\rm 2}} } \left(\sin \left(x_{U_{i} } \right)\right)\right)\left(\left| \psi _{+} \right\rangle  \right)-i\left(\frac{{\rm 1}}{\sqrt{{\rm 2}} } \left(\cos \left(x_{U_{i} } \right)\right)-\frac{{\rm 1}}{\sqrt{{\rm 2}} } \left(\sin \left(x_{U_{i} } \right)\right)\right)\left(\left| \phi _{+} \right\rangle  \right)\right).} \end{split}
\end{equation} 
 Thus, up to the global phase, both states are the same.

Therefore, the $\left| \varphi \left(T_{U_{i} } \right)\right\rangle  _{ABC} $ system state of $ABC$ at $T_{U_{i} } $ is yielded as 
\begin{equation} \label{eq19} 
\begin{split} {\left| \varphi \left(T_{U_{i} } \right)\right\rangle  _{ABC} =\frac{{\rm 1}}{\sqrt{{\rm 2}} } \left| \xi \left(T_{U_{i} } \right)\right\rangle  _{AB} \left| +\right\rangle  +\frac{{\rm 1}}{\sqrt{{\rm 2}} } \left| \xi \left(T_{U_{i} } \right)\right\rangle  _{AB} \left| -\right\rangle  ,} \end{split} 
\end{equation} 
therefore, the resulting time $AB$ state at $t=T_{U_{i} } =x_{U_{i} } +\left(\pi /{\rm 4}\right)$ and $x_{U_{i} } \ne {\rm 0}$, $\left| \xi \left(T_{U_{i} } \right)\right\rangle  _{AB} $ is a non-maximally entangled system 
\begin{equation} \label{eq20} 
\begin{split} {\left| \xi \left(T_{U_{i} } \right)\right\rangle  _{AB}} ={}& {\frac{{\rm 1}}{\sqrt{{\rm 2}} } } {\left(\cos \left(x_{U_{i} } \right)\right)+\frac{{\rm 1}}{\sqrt{{\rm 2}} } \left(\sin \left(x_{U_{i} } \right)\right)\left(\left| \psi _{+} \right\rangle  \right)} 
\\ -&{i\left(\frac{{\rm 1}}{\sqrt{{\rm 2}} } \left(\cos \left(x_{U_{i} } \right)\right)-\frac{{\rm 1}}{\sqrt{{\rm 2}} } \left(\sin \left(x_{U_{i} } \right)\right)\right)\left(\left| \phi _{+} \right\rangle  \right),} \end{split}
\end{equation} 
 with entanglement between user $U_{i} $ and $B_{i} $ as 
\begin{equation} \label{eq21} 
E^{\left(T_{U_{i} } \right)} \left(U_{i} :B_{i} \right)=\sin ^{2} \left(2\left(\frac{\pi }{4} +x_{U_{i} } \right)\right). 
\end{equation} 
 
\subsection{Differentiation in the Time Domain }

In the time domain differentiation service, a transmit user $U_{i} $ generates the initial system $ABC$, transmits separable $B$ to receiver $B_{i} $, and applies the local unitary $U_{AC} $ on subsystem $AC$ for time $T_{U_{i} } \left(\pi /{\rm 4}\right)$ (using the core protocol of Phase 2). Using the oscillation period $T_{\pi } \left(U_{i} :B_{i} \right)$ distributed in Phase 1, the resulting $AB$ subsystem after total time 
\begin{equation} \label{eq22} 
T=T_{U_{i} } \left(\pi /{\rm 4}\right)=T_{\pi } \left(U_{i} :B_{i} \right)/{\rm 4} 
\end{equation} 
 between users $U_{i} $ and $B_{i} $, $i=1,\ldots ,K$ is a maximally entangled system, $E^{\left(T_{\pi } \right)} \left(U_{i} :B_{i} \right)=1$, for all $i$.
  
\subsubsection{Differentiation Service}

 The Phases 1 and 2 of the time domain differentiation method (Protocol 2) are as included in Protocol 2. 

\begin{prot}
  \DontPrintSemicolon
\caption{Differentiation in Time Domain}
\textbf{Step 1}. Let $U_{i} ,i=1,\ldots ,K$ be the set of transmit users, and $B_{i} ,i=1,\ldots ,K$ are the receiver users. Let $T_{\pi } \left(U_{i} :B_{i} \right)$ be the oscillation time selected for user pairs $U_{i} $ and $B_{i} $, and let $T_{U_{i} } \left(\pi /{\rm 4}\right)$ be defined as 
\begin{equation} \label{eq23} 
T_{U_{i} } \left(\frac{\pi }{{\rm 4}} \right)=\frac{T_{\pi } \left(U_{i} :B_{i} \right)}{{\rm 4}} . 
\end{equation} 
 For all $i$, distribute the oscillation period of time $T_{\pi } \left(U_{i} :B_{i} \right)$ information via an encoder unit ${\mathcal{E}}$ between $U_{i} $ and $B_{i} $ (Phase 1).

\textbf{Step 2}. In the transmit user $U_{i} $, generate the initial system $ABC$, transmit separable B to receiver $B_{i} $, and apply the local unitary $U_{AC} $ on subsystem $AC$ for time $T_{U_{i} } \left(\pi /{\rm 4}\right)$ (Core protocol of Phase 2 between the users).

\textbf{Step 3}. The resulting $AB$ subsystem after total time $T=T_{U_{i} } \left(\pi /{\rm 4}\right)=T_{\pi } \left(U_{i} :B_{i} \right)/{\rm 4}$ between users $U_{i} $ and $B_{i} $ is a maximally entangled system, $E^{\left(T_{\pi } \right)} \left(U_{i} :B_{i} \right)=1$, for all $i$.

\end{prot}

\paragraph{Description}

Let us focus on a particular $ABC$ of users $U_{i} $ and $B_{i} $. The same results apply for all users of the network.

After the steps of Protocol 2, the density $\sigma _{ABC} $ of the final $ABC$ state is as 
\begin{equation} \label{eq24} 
\begin{split} {\sigma _{ABC} } {} &={\left| \varphi \left(t\right)\right\rangle  \left\langle \left. \varphi \left(t\right)\right|\right. _{ABC} =U\rho _{0} U^{\dag } } \\ {} & ={\frac{{\rm 1}}{{\rm 2}} \left(U_{AC} \left| \psi _{+} \right\rangle  \left\langle \left. \psi _{+} \right|\right. \left| +\right\rangle  \left\langle \left. +\right|\right. U_{AC}^{\dag } \right)+\frac{{\rm 1}}{{\rm 2}} \left(U_{AC} \left| \phi _{+} \right\rangle  \left\langle \left. \phi _{+} \right|\right. \left| -\right\rangle  \left\langle \left. -\right|\right. U_{AC}^{\dag } \right),} \end{split}
\end{equation} 
 where $\left| \varphi \left(t\right)\right\rangle  _{ABC} $ at $t$ is evaluated as 
\begin{equation} \label{eq25} 
\begin{split} & {\left| \varphi \left(t\right)\right\rangle  _{ABC} } \\  & {=\frac{{\rm 1}}{\sqrt{{\rm 2}} } \left(U_{AC} \left(\left| \psi _{+} \right\rangle  \left| +\right\rangle  \right)+U_{AC} \left(\left| \phi _{+} \right\rangle  \left| -\right\rangle  \right)\right)} \\ & {=\frac{{\rm 1}}{\sqrt{{\rm 2}} } \left(\cos \left(t\right)\right. \left(\frac{{\rm 1}}{{\rm 2}} \left(\left| {\rm 010}\right\rangle  +\left| {\rm 011}\right\rangle  +\left| {\rm 100}\right\rangle  +\left| {\rm 101}\right\rangle  \right)\right)+{\rm cos}\left(t\right)\left(\frac{{\rm 1}}{{\rm 2}} \left(\left| 000\right\rangle  -\left| {\rm 001}\right\rangle  +\left| {\rm 110}\right\rangle  -\left| {\rm 111}\right\rangle  \right)\right)} 
\\ {} & -{\left. i\sin \left(t\right)\left(\frac{{\rm 1}}{{\rm 2}} \left(\left| {\rm 111}\right\rangle  +\left| {\rm 110}\right\rangle  +\left| {\rm 001}\right\rangle  +\left| 000\right\rangle  \right)\right)+i\sin \left(t\right)\left(\frac{{\rm 1}}{{\rm 2}} \left(\left| {\rm 101}\right\rangle  -\left| {\rm 100}\right\rangle  +\left| {\rm 011}\right\rangle  -\left| {\rm 010}\right\rangle  \right)\right)\right)} \\ & {=\frac{{\rm 1}}{\sqrt{{\rm 2}} } \left(\cos \left(t\right)\left(\left| \psi _{+} \right\rangle  \left| +\right\rangle  +\left| \phi _{+} \right\rangle  \left| -\right\rangle  \right)-i\sin \left(t\right)\left(\left| \phi _{+} \right\rangle  \left| +\right\rangle  -\left| \psi _{+} \right\rangle  \left| -\right\rangle  \right)\right)} \\ & {=\frac{{\rm 1}}{\sqrt{{\rm 2}} } \left(\left(\cos \left(t\right)\left(\left| \psi _{+} \right\rangle  \right)-i\sin \left(t\right)\left(\left| \phi _{+} \right\rangle  \right)\right)\left| +\right\rangle  +\left(\cos \left(t\right)\left(\left| \phi _{+} \right\rangle  \right)+i\sin \left(t\right)\left(\left| \psi _{+} \right\rangle  \right)\right)\left| -\right\rangle  \right),} \end{split}
\end{equation} 
 where the sign change on $U_{AC} \left(\left| \phi _{+} \right\rangle  \left| -\right\rangle  \right)$ is due to the $\left| -\right\rangle  $ eigenstate on $C$.

Thus, at $t=\pi /{\rm 4}=T_{U_{i} } \left(\pi /{\rm 4}\right)$, the system state is 
\begin{equation} \label{eq26} 
\begin{split} & {\left| \varphi \left(T_{U_{i} } \left(\pi /{\rm 4}\right)\right)\right\rangle  _{ABC} } \\ {} & {=\frac{{\rm 1}}{\sqrt{{\rm 2}} } \left(\cos \left(\pi /{\rm 4}\right)\left(\left| \psi _{+} \right\rangle  \left| +\right\rangle  +\left| \phi _{+} \right\rangle  \left| -\right\rangle  \right)-i\sin \left(\pi /{\rm 4}\right)\left(\left| \phi _{+} \right\rangle  \left| +\right\rangle  -\left| \psi _{+} \right\rangle  \left| -\right\rangle  \right)\right)} \\ {} & {=\frac{{\rm 1}}{\sqrt{{\rm 2}} } \left(\frac{{\rm 1}}{\sqrt{{\rm 2}} } \left(\left| \psi _{+} \right\rangle  \left| +\right\rangle  +\left| \phi _{+} \right\rangle  \left| -\right\rangle  \right)-i\frac{{\rm 1}}{\sqrt{{\rm 2}} } \left(\left| \phi _{+} \right\rangle  \left| +\right\rangle  -\left| \psi _{+} \right\rangle  \left| -\right\rangle  \right)\right)} \\ {} & {=\frac{{\rm 1}}{\sqrt{{\rm 2}} } \left(\begin{array}{l} {\left(\frac{{\rm 1}}{\sqrt{{\rm 2}} } \left(\left| \psi _{+} \right\rangle  \right)-i\frac{{\rm 1}}{\sqrt{{\rm 2}} } \left(\left| \phi _{+} \right\rangle  \right)\right)\left| +\right\rangle  } \\ {+\left(\frac{{\rm 1}}{\sqrt{{\rm 2}} } \left(\left| \phi _{+} \right\rangle  \right)+i\frac{{\rm 1}}{\sqrt{{\rm 2}} } \left(\left| \psi _{+} \right\rangle  \right)\right)\left| -\right\rangle  } \end{array}\right),} \end{split}
\end{equation} 
 where 
\begin{equation} \label{eq27} 
\frac{{\rm 1}}{\sqrt{{\rm 2}} } \left(\left| \phi _{+} \right\rangle  +i\left| \psi _{+} \right\rangle  \right)=i\left(\frac{{\rm 1}}{\sqrt{{\rm 2}} } \left(\left| \psi _{+} \right\rangle  -i\left| \phi _{+} \right\rangle  \right)\right); 
\end{equation} 
 Thus, up to the global phase both states are the same yielding relative entropy of entanglement between users $U_{i} $ and $B_{i} $ as 
\begin{equation} \label{eq28} 
E^{\left(T_{\pi } \right)} \left(U_{i} :B_{i} \right)=1 
\end{equation} 
 with unit probability.

\subsection{Comparative Analysis}

\begin{center}
\begin{figure}[!h]
\begin{center}
\includegraphics[angle = 0,width=1\linewidth]{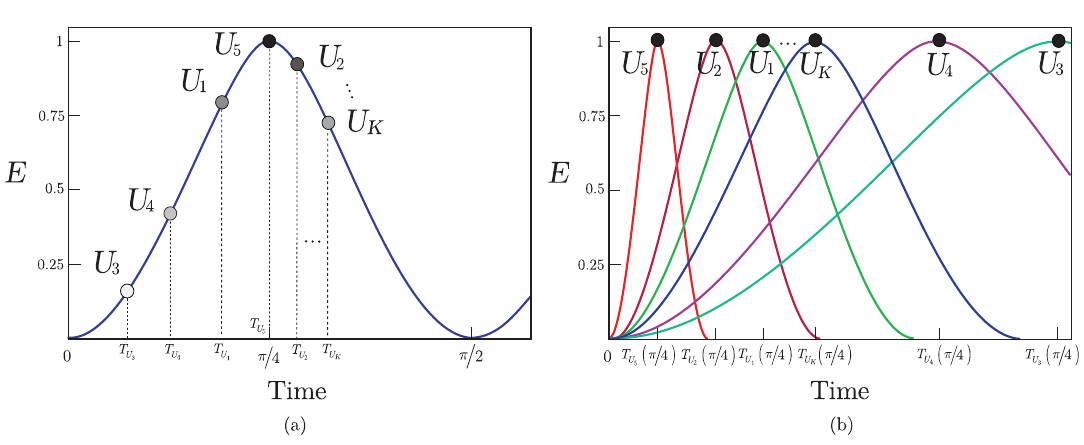}
\caption{Entanglement differentiation service via Hamiltonian dynamics in a multiuser environment. \textbf{(a):} Protocol 1. Each user gives a different amount of entanglement $E\left(U_{i} :B_{i} \right)\le {\rm 1}$ at a global period of time $T_{\pi } $. The differentiation is made in the amount of entanglement (relative entropy of entanglement) by applying the local unitaries for time $T_{U_{i} } $ for $U_{i} ,i=1,\ldots ,K$. User $U_{{\rm 5}} $ has the highest priority thus the user gets a maximally entangled system, user $U_{{\rm 3}} $ is the lowest priority user and associated with a low amount of entanglement. \textbf{(b):} Protocol 2. All users are assigned with a maximally entangled system, $E\left(U_{i} :B_{i} \right)=1$, and the differentiation is made in the time domain. For users $U_{i} ,B_{i} ,i=1,\ldots ,K$ a particular period of time $T_{\pi } \left(U_{i} :B_{i} \right)$ is assigned, and each local unitary is applied for $T_{U_{i} } \left(\pi /{\rm 4}\right)=T_{\pi } \left(U_{i} :B_{i} \right)/{\rm 4}$ time $t$o achieve maximally entangled states between the parties. User $U_{{\rm 5}} $ has the highest priority thus the user associated with the shortest time period, user $U_{{\rm 3}} $ is the lowest priority user with a long time period for the generation of a maximally entangled system.} 
 \label{fig2}
 \end{center}
\end{figure}
\end{center}

 The results of the proposed differentiation methods, Protocols 1 and 2, are compared in \fref{fig2}. \fref{fig2}(a) illustrates the results of a differentiation in the entanglement quantity, while \fref{fig2}(b) depicts the results of the time-domain differentiation method.

\section{Conclusions}
\label{sec4}
Entanglement differentiation is an important problem in quantum networks where the legal users have different priorities or where differentiation is a necessity for an arbitrary reason. In this work, we defined the EAD service for the availability of entanglement in quantum Internet. In EAD, the differentiation is either made in the amount of entanglement associated with a legal user or in the amount of time that is required to establish a maximally entangled system. The EAD method requires a classical phase for the distribution of timing information between the users. The entanglement establishment is based on Hamiltonian dynamics, which allows the efficient implementation of the entanglement differentiation methods via local unitary operations. The method requires no entanglement transmission between the parties, and the application time of the unitaries can be selected as arbitrarily small via the determination of the oscillation periods to achieve an efficient practical realization. The EAD method is particularly convenient for practical quantum networking scenarios, quantum communication networks, and future quantum Internet. 

\section*{Acknowledgements}
L.GY. would like to thank Tomasz Paterek for useful discussions. This work was partially supported by the National Research Development and Innovation Office of Hungary (Project No. 2017-1.2.1-NKP-2017-00001), by the Hungarian Scientific Research Fund - OTKA K-112125 and in part by the BME Artificial Intelligence FIKP grant of EMMI (BME FIKP-MI/SC).


\newpage
\appendix
\setcounter{table}{0}
\setcounter{figure}{0}
\setcounter{equation}{0}
\setcounter{algocf}{0}
\renewcommand{\thetable}{\Alph{section}.\arabic{table}}
\renewcommand{\thefigure}{\Alph{section}.\arabic{figure}}
\renewcommand{\theequation}{\Alph{section}.\arabic{equation}}
\renewcommand{\thealgocf}{\Alph{section}.\arabic{algocf}}

\section{Appendix}
\subsection{Steps of the Core Protocol}
\label{corep}
The detailed discussion of the Core Protocol (Protocol 0) is as follows. 

In Step 1, the input system $AB$ \eqref{eq1} is an even mixture of the Bell states which contains no entanglement. It is also the situation in Step 2 for the subsystem $AB$ of ${\rho }_{ABC}$ \eqref{eq3}, thus the relative entropy of entanglement for ${\rho }_{AB}$ is zero, $E\left(A:B\right)=0$.
The initial ${\rho }_{AB}$ in \eqref{eq1} and \eqref{eq3}, is the unentangled, Bell-diagonal state 
\begin{equation} \label{Aeq9} 
{{\rho }_{AB}}=\tfrac{1}{4}\left( \begin{matrix}
   1 & 0 & 0 & 1  \\
   0 & 1 & 1 & 0  \\
   0 & 1 & 1 & 0  \\
   1 & 0 & 0 & 1  \\
\end{matrix} \right)
\end{equation} 
with eigenvalues $v_+=\frac{1}{2},v_-=0,u_+=\frac{1}{2},u_-=0$.

In Step 3, dynamics generated by local Hamiltonian $H_{AC}={\sigma }^x_A{\sigma }^x_C$ with energy $E_{AC}$ will lead to entanglement oscillations in $AB$. Thus, if $U_{AC}$ is applied exactly only for a well determined time $t$, the local unitary will lead to maximally entangled $AB$ with a unit probability.

As a result, for subsystem $AB$, the entanglement $E\left(A:B\right)$ oscillates \cite{r10} with the application time $t$ of the unitary. In particular, the entanglement oscillation in $AB$ generated by the energy $E_{AC}$ \eqref{eq6} of the Hamiltonian $H_{AC}$ \eqref{eq5}. This oscillation has a period time $T_{\pi }$, which exactly equals to $4t$, thus 
\begin{equation} \label{Aeq10} 
T_{\pi }=4t, 
\end{equation} 
 where $t$ is determined by Alice and Bob. In other words, time $t$ identifies $\pi /4$, where $\pi $ is the oscillation period.

Therefore, after Step 3, the density ${\sigma }_{ABC}$ of the final $ABC$ state is as 
\begin{equation} \label{Aeq11} 
\begin{split}
{\sigma }_{ABC} & =\left|\varphi \left(t\right)\right\rangle {\left\langle \left.\varphi \left(t\right)\right|\right.}_{ABC}=U{\rho }_0U^{\dagger } \\ 
 & =\frac{1}{2}\left(U_{AC}\left|{\psi }_+\right\rangle \left\langle \left.{\psi }_+\right|\right.\left|+\right\rangle \left\langle \left.+\right|\right.U^{\dagger }_{AC}\right)+\frac{1}{2}\left(U_{AC}\left|{\phi }_+\right\rangle \left\langle \left.{\phi }_+\right|\right.\left|-\right\rangle \left\langle \left.-\right|\right.U^{\dagger }_{AC}\right), 
\end{split}
\end{equation} 
 where ${\left|\varphi \left(t\right)\right\rangle }_{ABC}$ at $t$ is evaluated as 
\begin{equation} \label{Aeq12} 
\begin{split}
 {\left|\varphi \left(t\right)\right\rangle }_{ABC}\\ 
  ={}&\frac{1}{\sqrt{2}}\left(U_{AC}\left(\left|{\psi }_+\right\rangle \left|+\right\rangle \right)+U_{AC}\left(\left|{\phi }_+\right\rangle \left|-\right\rangle \right)\right) \\ 
  ={}&\frac{1}{\sqrt{2}}\left(\cos \left(t\right)\right.\left(\frac{1}{2}\left(\left|010\right\rangle +\left|011\right\rangle +\left|100\right\rangle +\left|101\right\rangle \right)\right)+\cos \left(t\right)\left(\frac{1}{2}\left(\left|000\right\rangle -\left|001\right\rangle +\left|110\right\rangle -\left|111\right\rangle \right)\right) 
\\ 
 -& \left. i\sin \left(t\right)\left(\frac{1}{2}\left(\left|111\right\rangle +\left|110\right\rangle +\left|001\right\rangle +\left|000\right\rangle \right)\right)+i\sin \left(t\right)\left(\frac{1}{2}\left(\left|101\right\rangle -\left|100\right\rangle +\left|011\right\rangle -\left|010\right\rangle \right)\right)\right), \\ 
 &  
\end{split}
\end{equation} 
 that can be rewritten as 
\begin{equation} \label{Aeq13} 
\begin{split}
 & \frac{1}{\sqrt{2}}\left(\cos \left(t\right)\left(\left|{\psi }_+\right\rangle \left|+\right\rangle +\left|{\phi }_+\right\rangle \left|-\right\rangle \right)-i\sin \left(t\right)\left(\left|{\phi }_+\right\rangle \left|+\right\rangle -\left|{\psi }_+\right\rangle \left|-\right\rangle \right)\right) \\ 
 & =\frac{1}{\sqrt{2}}\left(\left(\cos \left(t\right)\left(\left|{\psi }_+\right\rangle \right)-i\sin \left(t\right)\left(\left|{\phi }_+\right\rangle \right)\right)\left|+\right\rangle +\left(\cos \left(t\right)\left(\left|{\phi }_+\right\rangle \right)+i\sin \left(t\right)\left(\left|{\psi }_+\right\rangle \right)\right)\left|-\right\rangle \right), 
\end{split}
\end{equation} 
 where the sign change on $U_{AC}\left(\left|{\phi }_+\right\rangle \left|-\right\rangle \right)$ is due to the $\left|-\right\rangle $ eigenstate on $C$.

Thus, at $t=\pi /4$, 
\begin{equation} \label{Aeq14} 
\begin{split}
{\left|\varphi \left(\pi /4\right)\right\rangle }_{ABC}&=\frac{1}{\sqrt{2}}\left(\cos \left(\pi /4\right)\left(\left|{\psi }_+\right\rangle \left|+\right\rangle +\left|{\phi }_+\right\rangle \left|-\right\rangle \right)-i\sin \left(\pi /4\right)\left(\left|{\phi }_+\right\rangle \left|+\right\rangle -\left|{\psi }_+\right\rangle \left|-\right\rangle \right)\right) \\ 
 & =\frac{1}{\sqrt{2}}\left(\frac{1}{\sqrt{2}}\left(\left|{\psi }_+\right\rangle \left|+\right\rangle +\left|{\phi }_+\right\rangle \left|-\right\rangle \right)-i\frac{1}{\sqrt{2}}\left(\left|{\phi }_+\right\rangle \left|+\right\rangle -\left|{\psi }_+\right\rangle \left|-\right\rangle \right)\right) \\ 
 & =\frac{1}{\sqrt{2}}\left( \begin{array}{l}
\left(\frac{1}{\sqrt{2}}\left(\left|{\psi }_+\right\rangle \right)-i\frac{1}{\sqrt{2}}\left(\left|{\phi }_+\right\rangle \right)\right)\left|+\right\rangle  \\ 
+\left(\frac{1}{\sqrt{2}}\left(\left|{\phi }_+\right\rangle \right)+i\frac{1}{\sqrt{2}}\left(\left|{\psi }_+\right\rangle \right)\right)\left|-\right\rangle  \end{array}
\right),
\end{split} 
\end{equation} 
 where 
\begin{equation} \label{Aeq15} 
\frac{1}{\sqrt{2}}\left(\left|{\phi }_+\right\rangle +i\left|{\psi }_+\right\rangle \right)=i\left(\frac{1}{\sqrt{2}}\left(\left|{\psi }_+\right\rangle -i\left|{\phi }_+\right\rangle \right)\right); 
\end{equation} 
i.e., up to the global phase both states are the same.

Therefore the ${\left|\varphi \left(\pi /4\right)\right\rangle }_{ABC}$ system state of $ABC$ at $t=\pi /4$ is yielded as 
\begin{equation} \label{Aeq16} 
{\left|\varphi \left(\pi /4\right)\right\rangle }_{ABC}=\frac{1}{\sqrt{2}}\left(\frac{1}{\sqrt{2}}\left(\left|{\psi }_+\right\rangle -i\left|{\phi }_+\right\rangle \right)\right)\left|+\right\rangle +\frac{1}{\sqrt{2}}\left(\frac{1}{\sqrt{2}}\left(\left|{\psi }_+\right\rangle -i\left|{\phi }_+\right\rangle \right)\right)\left|-\right\rangle , 
\end{equation} 
 while the density matrix ${\sigma }_{ABC}$ of the final $ABC$ system in matrix form is as
\begin{equation} \label{Aeq17} 
{{\sigma }_{ABC}}=\tfrac{1}{8}\left( \begin{matrix}
   1 & 0 & -i & 0 & -i & 0 & 1 & 0  \\
   0 & 1 & 0 & -i & 0 & -i & 0 & 1  \\
   i & 0 & 1 & 0 & 1 & 0 & i & 0  \\
   0 & i & 0 & 1 & 0 & 1 & 0 & i  \\
   i & 0 & 1 & 0 & 1 & 0 & i & 0  \\
   0 & i & 0 & 1 & 0 & 1 & 0 & i  \\
   1 & 0 & -i & 0 & -i & 0 & 1 & 0  \\
   0 & 1 & 0 & -i & 0 & -i & 0 & 1  \\
\end{matrix} \right).
\end{equation}

As one can verify, the resulting $AB$ state ${\left|\xi \left(\pi /4\right)\right\rangle }_{AB}$ at $t=\pi /4$ is pure and maximally entangled, 
\begin{equation} \label{Aeq18} 
\begin{split}
{\left|\xi \left(\pi /4\right)\right\rangle }_{AB}=\frac{1}{\sqrt{2}}\left(\left|{\psi }_+\right\rangle -i\left|{\phi }_+\right\rangle \right),
\end{split}
\end{equation} 
 yielding relative entropy of entanglement 
\begin{equation} \label{Aeq19} 
E\left(A:B\right)=1 
\end{equation} 
 with unit probability.

The ${\sigma }_{AB}$ density matrix of the final $AB$ state is 
\begin{equation} \label{Aeq20} 
\begin{split}
{\sigma }_{AB}&=\left|\xi (\pi /4)\right\rangle {\left\langle \xi (\pi /4)\right|}_{AB} \\ 
 & =\frac{1}{2}\left(\left|{\psi }_+\right\rangle -i\left|{\phi }_+\right\rangle \right)\left(\left\langle \left.{\psi }_+\right|\right.+i\left\langle \left.{\phi }_+\right|\right.\right) \\ 
 & =\frac{1}{2}\left(\left|{\psi }_+\right\rangle \left\langle \left.{\psi }_+\right|\right.+i\left|{\psi }_+\right\rangle \left\langle \left.{\phi }_+\right|\right.-i\left|{\phi }_+\right\rangle \left\langle \left.{\psi }_+\right|\right.+\left|{\phi }_+\right\rangle \left\langle \left.{\phi }_+\right|\right.\right), 
\end{split}
\end{equation} 
 which in matrix form is as 
\begin{equation} \label{Aeq21} 
{{\sigma }_{A}}_{B}=\tfrac{1}{4}\left( \begin{matrix}
   1 & -i & -i & 1  \\
   i & 1 & 1 & i  \\
   i & 1 & 1 & i  \\
   1 & -i & -i & 1  \\
\end{matrix} \right).
\end{equation} 
 The negativity for the ${\sigma }^{T_B}_{AB}$ partial transpose of ${{\sigma }_A}_B$ yields 
\begin{equation} \label{Aeq22} 
\text{N}\left({\sigma }^{T_B}_{AB}\right)=\tfrac{\left\|{\sigma }^{T_B}_{AB}\right\|-1}{2}=\tfrac{\text{Tr}\left(\sqrt{{\left({\sigma }^{T_B}_{AB}\right)}^{\dagger }{\sigma }^{T_B}_{AB}}\right)-1}{2}=\tfrac{i}{2}, 
\end{equation} 
 which also immediately proves that $AB$ is maximally entangled. For a comparison, for the density matrix of initial $AB$, \eqref{eq1}, is $\text{N}\left({\rho }^{T_B}_{AB}\right)=0$.

Note that subsystem $C$ requires no further storage in a quantum memory, since the output density ${\sigma }_{ABC}$ can be rewritten as 
\begin{equation} \label{Aeq23} 
\begin{split}
 {{\sigma }_{ABC}} &=\frac{1}{2}\left( {{\left| \xi \left( \pi /4 \right) \right\rangle }_{AB}}\left| + \right\rangle  \right)\left( {{\left\langle  \xi \left( \pi /4 \right) \right|}_{AB}}\left\langle  + \right| \right)+\frac{1}{2}\left( {{\left| \xi \left( \pi /4 \right) \right\rangle }_{AB}}\left| - \right\rangle  \right)\left( {{\left\langle  \xi \left( \pi /4 \right) \right|}_{AB}}\left\langle  - \right| \right) \\ 
 & =\left| \xi \left( \pi /4 \right) \right\rangle {{\left\langle  \xi \left( \pi /4 \right) \right|}_{AB}}\left( \left| + \right\rangle \left\langle  + \right|+\left| - \right\rangle \left\langle  - \right| \right) \\ 
 & =\left( \left| \xi \left( \pi /4 \right) \right\rangle {{\left\langle  \xi \left( \pi /4 \right) \right|}_{AB}} \right)I,
\end{split}
\end{equation} 
 where $I$ is the identity operator, therefore the protocol does not require long-lived quantum memories.

\subsubsection{Classical Correlations}
The classical correlation is transmitted subsystem $B$ of \eqref{eq1} in Step 1 is as follows. Since ${\rho }_{AB}$ is a Bell-diagonal state \cite{r49} of two qubits $A$ and $B$ it can be written as 
\begin{equation} \label{Aeq25} 
{\rho }_{AB}=\frac{1}{4}\left(I+\sum^3_{j=1}{}c_j{\sigma }^A_j\otimes {\sigma }^B_j\right)=\sum_{a,b}{}{\lambda }_{ab}\left|{\beta }_{ab}\right\rangle \left\langle \left.{\beta }_{ab}\right|\right., 
\end{equation} 
 where terms ${\sigma }_j$ refer to the Pauli operators, while $\left|{\beta }_{ab}\right\rangle $ is a Bell-state 
\begin{equation} \label{Aeq26} 
\left|{\beta }_{ab}\right\rangle =\frac{1}{\sqrt{2}}\left(\left|0,b\right\rangle +{\left(-1\right)}^a\left|1,1\oplus b\right\rangle \right), 
\end{equation} 
 while ${\lambda }_{ab}$ are the eigenvalues as 
\begin{equation} \label{Aeq27} 
{\lambda }_{ab}=\frac{1}{4}\left(1+{\left(-1\right)}^ac_1-{\left(-1\right)}^{a+b}c_2+{\left(-1\right)}^bc_3\right). 
\end{equation} 
 The $\mathcal{I}$ quantum mutual information of Bell diagonal state ${\rho }_{AB}$ quantifies the total correlations in the joint system ${\rho }_{AB}$ as 
\begin{equation} \label{Aeq28} 
\begin{split}
\mathcal{I} & =S\left({\rho }_A\right)+S\left({\rho }_B\right)-S\left({\rho }_{AB}\right) \\ 
 & =S\left({\rho }_B\right)-S\left(\left.B\right|A\right) \\ 
 & =2-S\left({\rho }_{AB}\right) \\ 
 & =\sum_{a,b}{}{\lambda }_{ab}\log _2\left(4{\lambda }_{ab}\right), 
\end{split}
 \end{equation} 
 where $S\left(\rho \right)=-\text{Tr}\left(\rho \log _2\rho \right)$ is the von Neumann entropy of $\rho $, and $S\left(\left.B\right|A\right)=S\left({\rho }_{AB}\right)-S\left({\rho }_A\right)$ is the conditional quantum entropy.

The $\mathcal{C}\left({\rho }_{AB}\right)$ classical correlation function measures the purely classical correlation in the joint state ${\rho }_{AB}$. The amount of purely classical correlation $\mathcal{C}\left({\rho }_{AB}\right)$ in ${\rho }_{AB}$ can be expressed as follows \cite{r49}: 
\begin{equation} \label{Aeq29} 
\begin{split}
 \mathcal{C}\left( {{\rho }_{AB}} \right)&=S\left( {{\rho }_{B}} \right)-\tilde{S}\left( \left. B \right|A \right) \\ 
 & =S\left( {{\rho }_{B}} \right)-\underset{{{E}_{k}}}{\mathop{\min }}\,\sum\limits_{k}{{{p}_{k}}S\left( {{\rho }_{\left. B \right|k}} \right)} \\ 
 & =1-H\left( \frac{1+c}{2} \right) \\ 
 & =\frac{1+c}{2}{{\log }_{2}}\left( 1+c \right)+\frac{1-c}{2}{{\log }_{2}}\left( 1-c \right),
\end{split} 
\end{equation} 
 where 
\begin{equation} \label{Aeq30} 
{{\rho }_{B|k}}=\frac{\langle  k |{{\rho }_{AB}}|k \rangle }{\langle  k | {{\rho }_{A}} |k \rangle }
\end{equation} 
 is the post-measurement state of ${\rho }_B$, the probability of result $k$ is 
\begin{equation} \label{Aeq31} 
{{p}_{k}}=D{{q}_{k}}\langle  k | {{\rho }_{A}}|k \rangle , 
\end{equation} 
 while $d$ is the dimension of system ${\rho }_A$ and the $q_k$ make up a normalized probability distribution, $E_k=Dq_k\left|k\right\rangle \left\langle \left.k\right|\right.$ are rank-one POVM (positive-operator valued measure) elements of the POVM measurement operator $E_k$ \cite{r49}, while $H\left(p\right)=-p\log _2p-\left(1-p\right)\log _2\left(1-p\right)$ is the binary entropy function, and 
\begin{equation} \label{Aeq32} 
c=\max \left|c_j\right|. 
\end{equation} 
 For the transmission of $B$ the subsystem ${\rho }_{AB}$ is expressed as given by \eqref{eq1}, thus the classical correlation during the transmission is 
\begin{equation} \label{Aeq33} 
\mathcal{C}\left({\rho }_{AB}\right)=1-H\left(\frac{1+c}{2}\right)=1, 
\end{equation} 
 where $c=1$.

\subsection{Abbreviations}
\begin{description}
\item[EAD] Entanglement Availability Differentiation
\item[POVM] Positive-Operator Valued Measure
\end{description}

\setlength{\arrayrulewidth}{0.1mm}
\setlength{\tabcolsep}{5pt}
\renewcommand{\arraystretch}{1.5}

\subsection{Notations}
The notations of the manuscript are summarized in \tref{tab1}.
\begin{center}
\begin{longtable}{||l|p{4.5in}||}
\caption{Summary of notations.}
\label{tab1}
\endfirsthead
\endhead
\hline
\textit{Notation} & \textit{Description} \\ \hline
$\rho _{ABC} $ & Initial system. \\ \hline 
$\sigma _{ABC} $ & Final system. \\ \hline 
$\rho _{AB} $, $\rho _{C} $ & Initial subsystems. \\ \hline 
${\left| \delta _{B}  \right\rangle} ^{\left(m,n\right)} $ & Subsystem $B$, ${\left| \varphi _{B}  \right\rangle} =\alpha {\left| 0 \right\rangle} +\beta {\left| 1 \right\rangle} $, encoded via an $\left(m,n\right)$ redundant quantum parity code as\newline ${\left| \delta _{B}  \right\rangle} ^{\left(m,n\right)} =\alpha {\left| \chi _{+}  \right\rangle} _{1}^{\left(m\right)} \ldots {\left| \chi _{+}  \right\rangle} _{n}^{\left(m\right)} +\beta {\left| \chi _{-}  \right\rangle} _{1}^{\left(m\right)} \ldots {\left| \chi _{-}  \right\rangle} _{n}^{\left(m\right)} $,\newline where ${\left| \chi _{\pm }  \right\rangle} ^{\left(m\right)} ={\left| 0 \right\rangle} ^{\otimes m} \pm {\left| 1 \right\rangle} ^{\otimes m} $.  \\ \hline 
$T$ & Period time selected by Alice and Bob.  \\ \hline 
${\rm {\mathcal N}}_{1\ldots n} $ & Intermediate quantum repeaters between Alice and Bob.  \\ \hline 
$\sigma ^{x} $ & Pauli X matrix. \\ \hline 
$H_{AC} $ & Hamiltonian, $H_{AC} =\sigma _{A}^{x} \sigma _{C}^{x} $. \\ \hline 
$E_{AC} $ & Energy of Hamiltonian $H_{AC} $. \\ \hline 
$U_{AC} $ & Unitary, applied by Alice on subsystem $AC$ for a time $t$, $U_{AC} =\exp \left(-iH_{AC} t\right)$, where $H_{AC} =\sigma _{A}^{x} \sigma _{C}^{x} $ is a Hamiltonian, $\sigma ^{x} $ is the Pauli X matrix. \\ \hline 
$t$ & Application time of unitary $U_{AC} $, determined by Alice and Bob.  \\ \hline 
$I$ & Identity operator. \\ \hline 
$\hbar $  & Reduced Planck constant. \\ \hline 
$E\left(\cdot \right)$ & Relative entropy of entanglement. \\ \hline 
$T_{\pi } $ & Oscillation period, $T_{\pi } =4t$,  where $\pi $ is the period. \\ \hline 
${\left| \xi \left({\textstyle\frac{\pi }{4}} \right) \right\rangle} _{AB} $ & Output $AB$ subsystem at time $t$,\newline ${\left| \xi \left({\textstyle\frac{\pi }{4}} \right) \right\rangle} _{AB} ={\textstyle\frac{1}{\sqrt{2} }} \left({\left| \psi _{+}  \right\rangle} -i{\left| \phi _{+}  \right\rangle} \right)$,\newline where ${\left| \psi _{+}  \right\rangle} ={\textstyle\frac{1}{\sqrt{2} }} \left({\left| 01 \right\rangle} +{\left| 10 \right\rangle} \right)$, ${\left| \phi _{+}  \right\rangle} ={\textstyle\frac{1}{\sqrt{2} }} \left({\left| 00 \right\rangle} +{\left| 11 \right\rangle} \right)$ are maximally entangled states. \\ \hline 
$\sigma _{AB}^{{{T}_{B}}}$ & Partial transpose of output AB subsystem ${{\sigma }_{A}}_{B}$. \\ \hline 
$N\left( \sigma _{AB}^{{{T}_{B}}} \right)$ & Negativity for the $\sigma _{AB}^{{{T}_{B}}}$ partial transpose of ${{\sigma }_{A}}_{B}$. \\ \hline 
\end{longtable}
\end{center}
\end{document}